\documentclass[sigconf]{acmart}

\usepackage{booktabs} 
\usepackage{graphicx}
\usepackage{color, colortbl}
\definecolor{Gray}{gray}{0.9}
\newcolumntype{g}{>{\columncolor{Gray}}p}
\definecolor{Black}{gray}{.7}
\newcolumntype{k}{>{\columncolor{Black}}p}

\def\BibTeX{{\rm B\kern-.05em{\sc i\kern-.025em b}\kern-.08em
    T\kern-.1667em\lower.7ex\hbox{E}\kern-.125emX}}
\usepackage{makecell}
\usepackage{url}

\renewcommand\footnotetextcopyrightpermission[1]{}
\begin{document}

\title{Linking Stakeholders' Viewpoint Concerns and Microservices-based Architecture
}

\author{ Mujahid Sultan$^{\dag,*}$ }
\affiliation{%
  \institution{$^{\dag}$ Department of Computer Science, Ryerson University, Toronto, Canada}
  \institution{$^*$ Treasury Board Secretariat, Government of Ontario, Toronto, Canada}
}
\email{mujahid.sultan@{ryerson.ca; ontario.ca}}

\acmDOI{xx.xxx/xxx_x}
\acmISBN{978-1-4503-8104-8/21/03}
\acmConference[SAC'21]{ACM SAC Conference}{March 22-March 26, 2021}{Gwangju, South Korea}
\acmYear{2021}
\copyrightyear{2021}
\acmPrice{15.00}

\begin{abstract}
Widespread adoption of \textit{agile project management}, \textit{independent delivery} with microservices, and \textit{automated deployment} with DevOps has tremendously speedup the systems development. The real game-changer is \textit{continuous integration (CI)}, \textit{continuous delivery} and \textit{continuous deployment (CD)}. Organizations can do multiple releases a day, shortening the test, release, and deployment cycles from weeks to minutes. 

Maturity of container technologies like Docker and container orchestration platforms like Kubernetes has promoted microservices architecture, especially in the cloud-native developments. Various tools are available for setting up CI/CD pipelines. Organizations are moving away from monolith applications and moving towards microservices-based architectures. Organizations can quickly accumulate hundreds of such microservices accessible via \textit{application programming interfaces} (APIs). 

The primary purpose of these modern methodologies is agility, speed, and reusability. While DevOps offers speed and time to market, agility and reusability may not be guaranteed unless microservices and API's are linked to enterprise-wide stakeholders' needs. The link between stakeholders' needs and microservices/APIs is not well captured nor adequately defined.

In this publication, we describe a structured method to create a logical link among APIs and microservices-based agile developments with enterprise stakeholders' needs and \textit{viewpoint concerns}. This method enables capturing and documenting enterprise-wide stakeholders' needs, whether these are business owners, planners (product owners, architects), designers (developers, DevOps engineers), or the partners and subscribers of an enterprise.
\end{abstract}


\maketitle

\section{Introduction}

The IT industry has seen many transformations in the Software Development Life Cycle (SDLC) methodologies and development approaches. SDLCs ranging from waterfall to agile, and the development approaches from monolith to microservices. The fundamental difference between the monolith and the microservices-based architectures is the codebase. Monolith applications work of a central and relatively large codebase, whereas microservices have a independent and relatively small codebase. Due to the agile SDLCs, new microservices can be designed very quickly and deployed in production immediately. Agile management practices and microservice designs culminated in a new way of developing software - the DevOps, where Continuous Integration (CI) and Continuous Delivery (CD) play a major role. Most of the medium to large organizations, like public and semi-public entities who remained married to structured methodologies (like enterprise architecture frameworks) of creating enterprise blueprints, find it very difficult to transition to this new software development approaches.

Business process management (BPM) and Enterprise Architecture (EA) disciplines emerged to streamline the business and systems gap. With the adoption of agile methodologies and DevOps, this gap is widening. 
EA management is a practice to document relationships among businesses and systems. Many EA frameworks have evolved \cite{pereira2004method}, and have received some maturity \cite{shah2007frameworks} over the past decade, TOGAF, FEAF, and Zachman, to name a few. A comprehensive review of EA and EA frameworks given in \cite{winter2006essential, bernard2012introduction, lankhorst2009enterprise}; and evaluating EA frameworks given in \cite{schekkerman2004survive, leist2006evaluation, tang2004comparative}. EA frameworks consist of artifacts, which are descriptions of the enterprise from a specific viewpoint\footnote{We use ANSI/IEEE Standard 1471-2000 definitions of stakeholders' views and viewpoints \cite{ieee2000recommended}. A view is a representation of a whole system from the perspective of a related set of concerns. A viewpoint defines the perspective from which a view is taken. A viewpoint is where you are looking from - the vantage point or perspective that determines what you see; a view is what you see.} of a group of stakeholders \cite{finkelstein1992viewpoints, rozanski2012software}. The stakeholders' groups are  owners, designers (architects), systems engineers, developers and DevOps. 

\begin{figure}[!ht]
	\centerline{
	{(a)}
	\includegraphics[width=39mm, height=35mm]{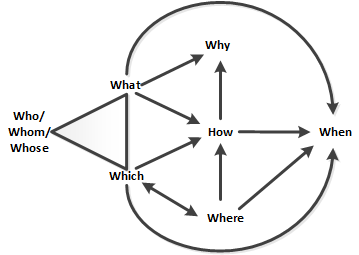} 
	{(b)}
	\includegraphics[width=39mm, height=35mm]{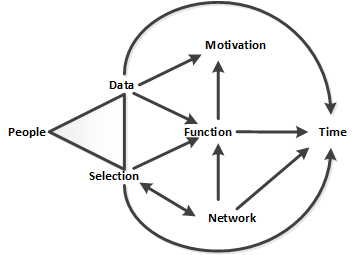}
	}
	\caption{(a) Order and inter-dependencies of English Language Interrogatives based on \protect \cite{cysouw2004interrogative} and \protect \cite{koenig2004any}. (b) In System terms, the material categories corresponding to the interrogative words. \textit{Legend:} The edges represent relationships among interrogatives, and the arrows point to dependent interrogatives. Directionless edges indicate an absence of dependence among interrogatives. Bidirectional arrows indicate interrogatives are interchangeable and have strong dependency among each other.}
	\label{fig:one}
\end{figure}

John Zachman introduced the concept of Information System Architecture (ISA) in 1987 \cite{zachman1987framework}. The Zachman framework describes stakeholders' views focusing on five ``wh'' interrogatives (\emph{what}, \emph{who}, \emph{where}, \emph{why}, and \emph{when}) and one ``h'' interrogative (\emph{how}). This focus comes from journalism's w5h theory \cite{flint1917newspaper}. \citet{sultan2015ordering, sultan2018ordering} applied linguistic findings to establish that the w5h set of interrogatives is not complete and added interrogative \textit{which}. They denoted this new set of seven interrogatives as w6h. 
They demonstrated that asking questions, based on the w6h (in the order of precedence described in Figure~\ref{fig:one}), improves information flow for the description of stakeholders' \textit{viewpoint concerns}. 

While EA frameworks for monolith applications have matured, there is no structured approach for defining EA for microservices-based developments. In this publication, we extend the w6h EA framework and w6h requirements engineering pattern developed by \cite{sultan2015ordering, sultan2018ordering} to the microservices-based developments.

\textbf{Roadmap:} The rest of the paper is organized as follows. In Section~\ref{lit}, we explore the literature on defining EA for microservices-based developments. In Section~\ref{background}, we describe the technologies which enable microservice architectures. This section level sets the ground for the proposed method - given in Section~\ref{framework}. In Section~\ref{discussion}, we discuss the proposed framework and its suitability in the creation of EA framework for microservice-based developments. In Section~\ref{future} we give some future research directions.

\section{Related work} \label{lit}

\citet{thones2015microservices} and \citet{nadareishvili2016microservice} provided background on microservices. \citet{garriga2017towards} provided a complete analysis of microservices taxonomy, life cycle, and fitness in an organization. Benchmark requirements for microservices architecture research are given by \cite{aderaldo2017benchmark}. Challenges in Documenting microservice-based IT landscape is captured in the survey paper by \citet{kleehaus2019challenges} and \citet{soldani2018pains}. 


Reflections on SOA and microservices are presented by \citet{xiao2016reflections}. Contextual understanding of microservice architecture \textquotedblleft current and future directions" is given by \citet{cerny2018contextual}. Microservices architecture enablement with  DevOps are investigated by \citet{balalaie2016microservices}. Architectural patterns for microservices are given by \cite{taibi2018architectural}. 

There have been reasonable efforts and systematic approaches to identifying microservices from a monolith's system functional requirements  \cite{tyszberowicz2018identifying}. Several useful techniques for extracting microservice from monolith enterprise systems are given by \cite{levcovitz2016towards}.  \citet{mazlami2017extraction} proposed a formal approach to extract microservices from monolith applications based on the class structure but  did not consider stakeholders' \textit{viewpoint concerns}. \citet{o2017continuous} presented background on CD and CI and the DevOps enabling technologies but did not talk about the link between EA and microservices.

\citet{kratzke2016clouns} presented the anatomy of the cloud-native stack which is a good summary of components of cloud-native developments but it does not address the full set of enterprise stakeholders. A reference architecture for designing microservices is given by \citet{yu2016microservice}. They presented a good representation of the microservice-based reference architecture model but kept themselves at the technology view level.  \citet{bogner2016towards} investigated mechanisms for integrating microservice architecture and EA and presented a very high-level meta-model without getting into the details of the stakeholders' \textit{viewpoint concerns}.

None of these studies considered alignment between enterprise stakeholders' \textit{viewpoint concerns} and microservices. Our work gives a structured approach to address these concerns systematically.

The need to define a linkage between enterprise stakeholders' needs and microservices is well established in the literature. \citet{canat2018enterprise} in their recent paper titled \textquotedblleft Agile developments foe or friends with EA" found that 1) agile development and enterprise architecture can be combined, 2) there are clear communication problems among architects, different teams, and project owners, and 3) there is a lack of system and application reusability. 

Our contribution in this work is to address these issues. We propose a framework, a structured approach, and a mind-map to capture enterprise stakeholders' \textit{viewpoint concerns} for microservice-based developments. This framework can be used for analysis, re-use and strategic planning, filling the gap between enterprise architects and DevOps.

\section{Background} \label{background}

\begin{figure}[tb]
	\centerline{
	\includegraphics[width=70mm]{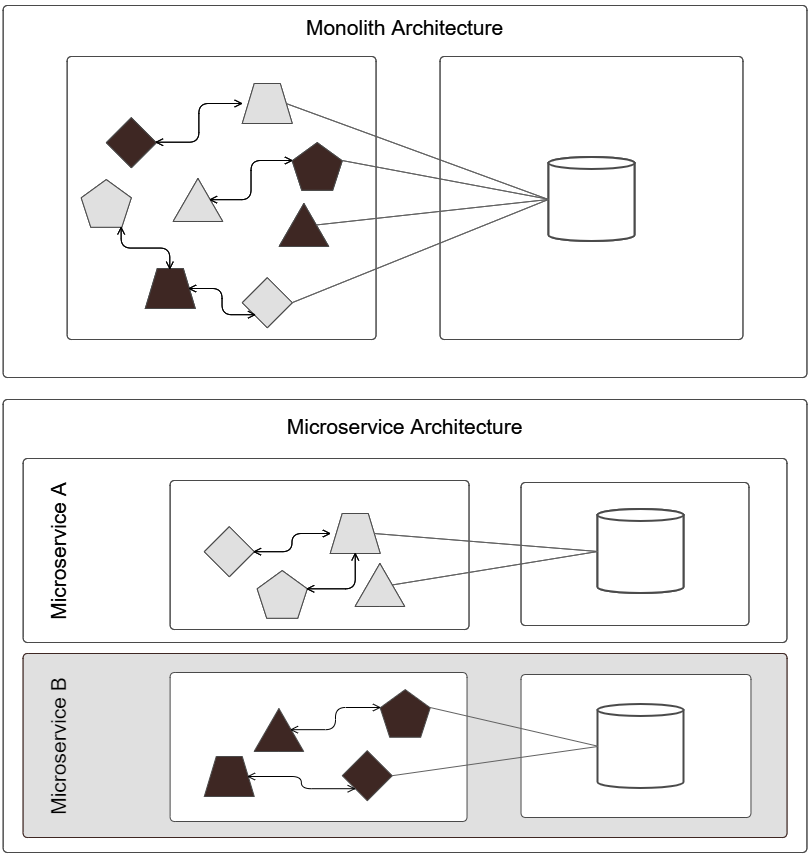} 
	}
	\caption{Monolith vs. Microservice Architecture. (Top) In the monolith architecture, there is a single source of persistent data for the entire application, mostly relational and normalized. (Bottom) Whereas, in the microservices architecture each microservice has its separate data store and is de-normalized.}
	\label{fig:microservices}
\end{figure}

\begin{figure}[tb]
	\centerline{
	{(a)}
	\includegraphics[width=35mm]{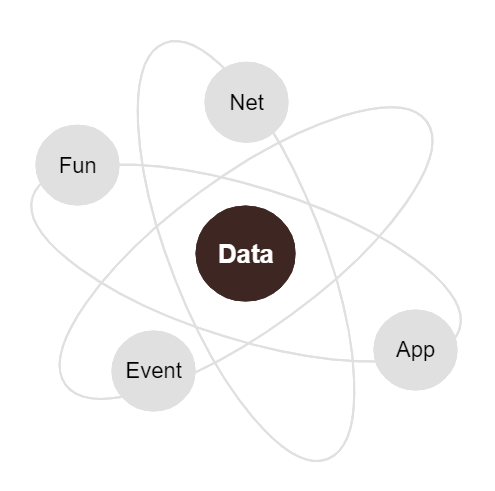} 
	{(b)}
	\includegraphics[width=35mm]{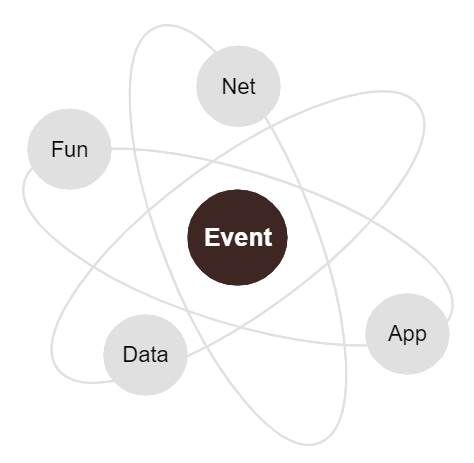}
	}
	\caption{(A) Monolith architecture: The data is the center of development, and is preserved first. Applications and systems depend upon data. (b) Microservices architecture: The Event is at the center, the data depends upon events.}
	\label{fig:enventcentric}
\end{figure}

The primary advantage, microservice-based architecture (as compared to monolith) offers, is independent development (in a language of choice), independent deployment, and loose coupling or access via the APIs. This paradigm warrants microservices to have their own piece of the data. The significant difference between monolith applications and microservices-based systems is shown in Fig. \ref{fig:microservices}. Similarly, this new way of designing systems puts the \textit{events} at the center compared to data-centric monolith application architectures. As shown in Fig~\ref{fig:enventcentric}, the enterprise's source of truth becomes events rather than a central repository.



Microservices are directly related to the business taxonomy of an enterprise and enable business process automation. APIs open an organizations' business processes for automation within the organization as well as for the business partners. 

APIs carry business value and the way the rest of the world interacts with an organization and are directly related to the  enterprise stakeholders' \textit{viewpoint concerns}. Microservices, on the other hand, automate business functions, sponsored by business stakeholders. Therefore, exploring the stakeholders' \textit{viewpoint concerns} for designing microservices and APIs is in order.

Microservices are discrete components of a system function or act as a composite application automating one or more business processes. Microservices are stand-alone applications automating business processes. One API can invoke many microservices to deliver a business function representing some part of the business process; this means the APIs can serve as a \textquotedblleft contract'' between two agents - providing the data or services (the output) in exchange for valid input. These attributes make APIs a valuable tool for business process automation.

To understand the link between Microservices and the Stakeholders' \textit{viewpoint concerns}, we briefly describe the enabling technologies which lead to the development of microservices in Section~\ref{cnap} through Section~\ref{paterns}.

A brief description of microservices, containers, and orchestration mechanisms is given below to put the stakeholders' viewpoints in the perspective of DevOps (where the developers are the ones who develop and put things in operation) and cloud-native developments.

\subsection{Cloud-Native Application Platforms}\label{cnap}
With the advent of cloud computing (on-demand availability of computer resources without direct active management by the user), most of the compute services can be offered as a service, for example, infrastructure as a service (IaaS), platform as a service (PaaS) and software as a service (SaaS). The microservices revolution happened with the advent of cloud-native developments on IaaS, which means that the organizations can develop their applications with their own data; the provider manages infrastructure only.  


\subsection{Microservices}
A microservice is a complete application in itself, full filling an entire business function, with its own development methodology and technology stack.  Microservices are attributed as radially available and on-demand consumable. Not all microservices are accessible to the outside world; these are called system or interface microservices. Similarly, some microservices are designed for synchronization or cleaning and are not discoverable or consumable on demand. Therefore, we can classify microservices in three broad categories, a) presentation-level, b) system-level, c) and integrity, or consistency check microservices. Each microservice has a different set of stakeholders' groups associated with it.  Therefore, capturing and aligning their \textit{viewpoint concerns} is essential for the enterprise.

The most common access mechanism for microservices is API endpoints. The endpoints need to have a registry for internal as well as external use. Externally these are used via the reverse proxy, whereas internally via some discovery and orchestration mechanism. The microservice architecture style's primary purpose is reusability, necessitating the need to align and link these APIs with the stakeholders' \textit{viewpoint concerns}. Different stakeholders pay for different microservices in an enterprise. Therefore, their needs should be accounted for, in the design of the microservices. 

One of the core principles of the microservice developments is independent deployment. Independent deployability allows on-demand scaling.  Suppose some part of the system experiences high load;  in this case, we can re-deploy or move a microservice to an environment with more resources. Without having to scale up hardware capacity for the entire enterprise system, this became possible with the maturity of the container technologies - briefly described below.  

\subsection{Containers}

Though containers have been around for a long time from \textit{Linux}, \textit{Solaris}~\cite{lageman2005solaris}, \textit{LXC}~\cite{bernstein2014containers}, \textit{Virtuozzo}~\cite{virtuozzo} to \textit{Docker}, the real change was brought by Docker containers \cite{bernstein2014containers}. Docker \cite{merkel2014docker} isolated resources that enable the packaging of applications with all dependencies installed, allowing the apps to run anywhere. Docker containers offer portability, performance, agility, isolation, and scalability of the applications. Despite the technical superiority, the platform's popularity is mainly due to the image registry (image of a container), where creators can create images, and the consumers can download and use, called Docker-hub. The registry can be public as well as private. Once the organization creates an image, it can be stored in the repository and pulled when and where needed - making this a convenient feature in a microservice architecture.

\subsection{Container Orchestration} Once containers are created, these can be deployed anywhere independently; the need arises to orchestrate the deployment of these containers on different infrastructure segments. Many products have emerged in this space, for example Docker Swarm~\cite{dockerdocumentation_2020}, Kubernetes \cite{kubernetes}, Apache Mesos~\cite{apachemesos}, and Nomad~\cite{nomadbyhashicorp}. However, Kubernetes have gained the most popularity and is widely used. Kubernetes (K8s) is an open-source system for automating deployment, scaling, and managing containerized applications. The docker containers are hosted in \textit{pods} as a basic unit of computations in K8s. The major components of Kubernetes are a master node and a worker node. Worker nodes run a Docker engine and can host multiple containers. An agent called \textit{Kubelet} runs at each worker node, and the \textit{Kube-proxy} handles each node networking. The \textit{container-runtime} is responsible for running the containers on each worker. The master node components are \textit{Kube-apiserver} which is a control plane and controls entire workers and \textit{pods}. \textit{Kube-schedular} schedules \textit{pods} and \textit{Kube-controller-manager} performs cluster operations and a distributed key-value store called \textit{Etcd}, which stores the data and maintains states.

\subsection{Service Mesh} 
When the number of microservices grows in number and complexity, the interaction between these becomes challenging - the discovery, load balancing, failure recovery, metrics, and rate-limiting, etc. The \textit{Service Mesh} - is a networking component that sits on top of the container orchestration layer, and takes care of these issues with custom control planes e.g., Istio~\cite{istio}, Linkerd~\cite{linkerd} and Conduit~\cite{conduit}, etc. The main components of service mesh are a \textit{Data plane} - which is a lightweight proxy as \textquotedblleft side-car"\footnote{explained in Section~\ref{sidecar}}; and a \textit{Control plane} - which controls configurations of the proxies based on performance metrics. 

\subsection{Continuous Integration (CI),
Continuous Delivery,
Continuous Deployment (CD)}

CI is a new software engineering practice to speed up software delivery by decreasing integration times. Integrations are verified by an automated build and test \cite{fowler2006continuous}. Similarly, Continuous Delivery is an automated build with each check-in of the new code. And Continuous Deployment is a process of releasing changes from the repository to production. CI/CD pipeline is performed via automated tools like Ansible~\cite{ansible}, Drone~\cite{drone}, Spinnaker~\cite{spinnaker}, AWS CodePipeLine~\cite{awscodepipeline} and Otter~\cite{inedo}, etc.


\subsection{Complexity and Patterns}\label{paterns}
As shown in Fig. \ref{fig:enventcentric} and Fig. \ref{fig:microservices} each microservice has its own copy of the data and the events become a source of truth for the data aggregation. Individual microservices can accumulate and modify their own data, but in case of dependency (just like the referential integrity in RDBMS) there are several ways to handle it, and the industry is still evolving and experimenting with different things.  The two most prominent patterns to address these issues are briefly described below.

\subsubsection{Event Sourcing Pattern}
One way to handle the data consistency in a microservices architecture via event sourcing. Event sourcing persists the state of each microservice data as a sequence of state-changing events. When the state of a microservice data changes, a new event is appended to the list of events. Thus by persisting events in an event store becomes a single source of truth for the enterprise. 

\subsubsection{Side-car Pattern} \label{sidecar}
Side-car pattern name comes from the side-car with a motorcycle, with the same life cycle as the microservice, and is tasked to perform peripheral tasks, such as monitoring, logging and, data integrity checks. Airbnb's SmartStack~\cite{airbnbeng_2016} and Netflix Prana~\cite{netflix} are some examples.



\section{Methods} \label{methods}
To describe and capture the stakeholders' \textit{viewpoint concerns}, from microservices perspective, we take Zachman's definition of stakeholders' views as given below:
\begin{enumerate}
	\item [A)] Scope (Ballpark View),
	\item [B)] Business Model (Owner's View),
	\item [C)] System Model (Designer's View),
	\item [D)] Technology Model (Builder's View),
	\item [E)] Detailed Representations (Subcontractor's View),
\end{enumerate}

In the following, we define each view and the group of stakeholders who hold this view. Then we briefly describe  stakeholders' \textit{viewpoint concerns} from each of w6h viewpoints of an enterprise for microservice-based developments. These descriptions constitute EA from a microservices perspective in the order given by \citet{sultan2018ordering}.

\subsection{Scope (Ballpark View)}\label{ballpark}		\textquotedblleft Ballpark View\textquotedblright\ sets the scope and puts architecture effort in perspective, and is also called the `contextual' view of the organization and describes the environment and surroundings the enterprise will function in and interact with.
The following groups of stakeholders see the organization from this view: business development directors, delivery managers, CIOs, CFOs, and CSOs, etc. Below is the detailed description of their \textit{viewpoint concerns} from each viewpoint perspective. 

These stakeholders need to capture the contextual information of APIs and Microservices (to be provisioned by the organization).

\begin{description}
    \item [\textit{who (people):}] 
    \item
            This stakeholders' \textit{viewpoint concern} captures the list of businesses an enterprise interacts with and may expose microservices to. The list of the organizations an enterprise might interchange information via APIs. This \textit{viewpoint concern} is used in planning for the business to business (B2B) and business to consumer (B2C) relationships.
    \item [\textit{what (data):}]
    \item
            This stakeholders'\textit{viewpoint concern} captures a list of things an enterprise needs to conduct business with other businesses and organizations - used in the design of microservices and may be required to offer  APIs.
    \item [\textit{which (selection):}]
    \item
            In this \textit{viewpoint concern}, we need to capture and select the list of organizations important to the enterprise - the target audiences of microservices and APIs important to business or vice versa - select microservices and APIs to conduct business.
    \item [\textit{where (network):}]
    \item
            This \textit{viewpoint concern} captures locations where the enterprise conducts its business - the candidate locations for microservices. The list of partner locations is also of interest form this viewpoint for planning purposes.
	\item [\textit{how (function):}]
    \item
        It is essential to define business processes required for microservices and APIs at the contextual level. This allows for strategic planning and setting future goals at the highest level of the enterprise hierarchy.

    \item [\textit{why (motivation):}]
    \item
           It becomes important to link the vision and business reasons for the provision of microservices and APIs at the contextual level. Without this linkage, the IT may provide services and offer APIs without business need.
    \item [\textit{when (time):}]
    \item
            Only the enterprise owners and planners can think and provide guidance on the business cycles important to the business: the target cycle of microservices and APIs renewal, retirement, or up-gradation is captured in this \textit{viewpoint concern}.
\end{description}

Capturing these \textit{viewpoint descriptions} from the  \emph{microservices} perspective adds to the enterprise's contextual view and enables planning for DevOps, re-user, vision and strategic direction.

\subsection{Business Model (Owner's View)} \textquotedblleft Business Model\textquotedblright\ describes the owner's view of the enterprise. What the enterprise does and why is captured in the \textquotedblleft business plan\textquotedblright. A business plan describes the value proposition, customer segmentation, and the channels business wants to reach to the customers. \textquotedblleft For whom are we creating the value\textquotedblright\ and \textquotedblleft which ones of our customer\textquoteright s problems\textquotedblright\ we are helping to solve, are the usual value propositions. 

The following groups of stakeholders see the enterprise from this view: shareholders, investors, founders, the board of governors, etc. 
Below is the detailed description of their \textit{viewpoint concerns} from each viewpoint perspective.

These stakeholders need to look at the conceptual level of the enterprise to see what API and Microservices will be provisioned by the organization, when, and why.

\begin{description}
    \item [\textit{who (people):}] 
    \item
        This \textit{viewpoint concern} identifies workflows required to expose microservices to internal or external audiences (organizations) via APIs
    \item [\textit{what (data):}]
    \item
            This \textit{viewpoint concern} describes the data elements (at semantic level) required for the design of microservices and APIs 
    \item [\textit{which (selection):}]
    \item
            This \textit{viewpoint concern} deals with prioritization and, selection of the important business processes, business scenarios, business functions required to provide APIs and microservices.
    \item [\textit{where (network):}]
    \item
        This \textit{viewpoint concern} captures locations (geographic locations) where specific microservices may reside (for example, the data center location for latency or specific service offering), and APIs may be offered.
	\item [\textit{how (function):}]
    \item
            This \textit{viewpoint concern} captures and defines at the business function level, which business functions will be automated with what microservice. Similarly, this \textit{viewpoint concern} also defines which business functions will be exposed via APIs. Table~\ref{owner_function} below captures this \textit{viewpoint concern} in tabular form.
            
            \begin{table}[H]\caption{business taxonomy offered via $\mu$-services and APIs} \label{owner_function}
            \centering
            \begin{tabular}{|c|c|c|} 
                 \hline
                  \rowcolor{Black}
                    \cellcolor{Black}\textbf{Business Taxonomy } &
                    \textbf{$\mu$-service$_{1} .. _{n}$} &
                	\textbf{API$_{1} .. _{n}$}  \cr
                \hline    
                    \cellcolor{Black} Business Function$_{1}$ & 
                    x& 
                    {x} \cr
                \hline 
                  \cellcolor{Black}Business Function$_{n}$ &
                    & 
                    x \cr
                \hline
            \end{tabular}
            \end{table}
    \item [\textit{why (motivation):}]
    \item
        It is imperative to capture and define the motivation and reasons for the provision of essential business services which are the target for microservices and APIs. This \textit{viewpoint concern} captures and documents these.
    \item [\textit{when (time):}]
    \item

        This \textit{viewpoint concern} captures business cycles for which a specific API or microservice is provided, modified, updated, or expired.

\end{description}

\subsection{System Model (Designer's View)}
The systems model is the view of an enterprise from the system designer's perspective (automation perspective). Where the business processes (candidates for automation) are described in microservice terms. At this stage, the designers (system engineers/architects) identify business concepts (entities) on which the microservices will work (use-cases, application components, etc.) or the entities used by microservices and APIs. API naming conventions and naming guidelines are documented. The CD and CI pipelines are decided, and tools are selected.

The following groups of stakeholders see the organization from this view: enterprise architects, requirements engineers, project managers, security architects, privacy specialists, internal and external regulators, auditors (internal and external), business continuity \& disaster recovery planners, etc. Below is the detailed description of their \textit{viewpoint concerns} from each viewpoint perspective. 

\begin{description}
    \item [\textit{who (people):}] 
    \item
        This \textit{viewpoint concern} captures interface requirements for APIs and microservices. A Decision which microservices will be system-level microservices and which ones will be user-level microservice is decided in this \textit{viewpoint concern}. Reverse proxy requirements and considerations from the access perspective may also be captured in this \textit{viewpoint concern}. This can be captured in a table like a grid, as shown in Table~\ref{b2bb2c}.
	   
            \begin{table}[H] 
                \caption{Interface requirements for $\mu$-services and APIs}
                \label{b2bb2c}
            \centering
            \begin{tabular}{|c|c|c|c|} 
                 \hline
                  \rowcolor{Black}
                    \textbf{Organization $\rightarrow$} &
                    \textbf{internal / external} &
                	\textbf{B2B} &
                	\textbf{B2C} \cr
                \hline    
                    \textbf{$\mu$-service$_{1} .. _{n}$} & 
                    x& 
                    {x}& \cr
                \hline 
                    \textbf{API$_{1} .. _{n}$} &
                    x& 
                    & \cr
                \hline
                  \textbf{reverse proxy} &
                    & 
                    x& 
                    x \cr
                \hline
            \end{tabular}
            \end{table}
    
    \item [\textit{what (data):}]
    \item
            This \textit{viewpoint concern} captures logical data elements to be used by each microservice and exposed by APIs. This is the point the system designers may decide the microservice patterns (\textquotedblleft event sourcing," \textquotedblleft side-car" pattern, etc. \cite{taibi2018architectural}) to be used for persistent data, as shown in Table~\ref{sys_data}.
        
            \begin{table}[H]
                \caption{Persistence requirements of $\mu$-services}
                \label{sys_data}
                \centering
            \begin{tabular}{|c|c|c|} 
                 \hline
                  \rowcolor{Black}
                    \textbf{Data elements} &
                    \textbf{Pattern} &
                	\textbf{Persistence} \cr
                \hline    
                    \textbf{$\mu$-service$_{1} .. _{n}$} & 
                    {side-car}& x\cr
                \hline 
                    \textbf{API$_{1} .. _{n}$} &
                    event-sourcing& \cr
                \hline
            \end{tabular}
            \end{table}
            
    \item [\textit{which (selection):}]
    \item
        Selection and prioritization based on the business value is captured in this \textit{viewpoint concern} - Which business processes, business scenarios, business functions and logical data elements are required for which APIs and microservices. Please note that \textit{which} quantifies the selection as described in detail by \cite{sultan2015ordering}.
    \item [\textit{where (network):}]
    \item
        This \textit{viewpoint concern} captures decisions for the provision of microservice at different network locations. The decisions to group services in different Kubernetes pods and clusters are captured in this \textit{viewpoint concern}. This can be represented in tabular form as shown in Table~\ref{sys_netwrok} below. The API Gateway rules are also defined by this \textit{viewpoint concern}, as shown in Table~\ref{sys_apigateway}.

            \begin{table}[H]
                \caption{ $\mu$-services deployment schemes and polices}
                \label{sys_netwrok}
                \centering
            \begin{tabular}{|c|c|c|c|} 
                 \hline
                  \rowcolor{Black}
                    \textbf{Deployment} &
                    \textbf{POD$_{1} .. _{n}$} &
                	\textbf{Network$_{1} .. _{n}$} &
                	\textbf{Service Mesh}\cr
                \hline    
                    \textbf{$\mu$-service$_{1} .. _{n}$} & 
                    .yaml files& 
                    & 
                    .yaml files\cr
                \hline 
                  \textbf{Container$_{1} .. _{n}$} &
                    .yaml files& 
                    x& 
                    .yaml files\cr
                \hline
            \end{tabular}
            \end{table}

            \begin{table}[H]
                \caption{API design and reverse proxy rules}
                \label{sys_apigateway}
                \centering
            \begin{tabular}{|c|c|c|c|} 
                 \hline
                  \rowcolor{Black}\textbf{APIs} & 
                    \textbf{Design} &
                    \textbf{API Gateway rules} &
                	\textbf{endpoint to IP} \cr

                \hline    
                    \textbf{API$_{1}$} & 
                    .yaml/jason& 
                    rules& 
                    mappings\cr
                \hline 
                  \textbf{API$_{n}$} &
                    .yaml/jason& 
                    rules& 
                    mappings\cr
                \hline
            \end{tabular}
            \end{table}
	\item [\textit{how (function):}]
    \item
     This \textit{viewpoint concern} captures business process models for  provisioning APIs and microservices. As each microservice can be designed using a different technology stack, this can be captured as a grid, as shown in Table~\ref{sys_fun1} below. Similarly, the design parameters -  the output, the input and success criteria, and error methods along with the tools used to design the APIs can be captured in this \textit{viewpoint concern} as shown in Table~\ref{sys_fun2}
            \begin{table}[H]
                \caption{ $\mu$-services technology stack}
                \label{sys_fun1}
                \centering
            \begin{tabular}{|c|c|c|c|} 
                 \hline
                  \rowcolor{Black}
                    \textbf{Tech Stack} &
                    \textbf{Node} &
                	\textbf{Python} &
                	\textbf{GO} \cr
                \hline    
                    \textbf{$\mu$-service 1} & 
                    x& 
                    {x}& \cr
                \hline 
                \textbf{$\mu$-service 2} & 
                    x& 
                    {x}& \cr
                \hline 
                  \textbf{$\mu$-service n} &
                    & 
                    x& 
                    x \cr
                \hline
            \end{tabular}
            \end{table}

            \begin{table}[H]
                \caption{ $\mu$-services and API design details (CRUD/REST methods, success/error codes and design tools/standards)}
                \label{sys_fun2}
                \centering
            \begin{tabular}{|c|c|c|c|} 
                 \hline
                  \rowcolor{Black}
                    \textbf{API messages} &
                    \textbf{Method} &
                    \textbf{Tech} &
                	\textbf{Code}  \cr
                \hline    
                    \textbf{API 1} & 
                    get, post &
                    OpenAPI\cite{OpenAPI} & 
                    300 \cr
                \hline 
                \textbf{API 2} & 
                   delete &
                    {RAML\cite{RAML}} & 
                    201 \cr
                \hline 
                  \textbf{API n} &
                  update &
                    Swagger\cite{swagger} & 
                    200 \cr
                \hline
            \end{tabular}
            \end{table}

    \item [\textit{why (motivation):}]
    \item
          This \textit{viewpoint concern} captures business rules specific to microservices and APIs. And can be represented in tabular form, or a business rules engine can be used.
          
    \item [\textit{when (time):}]
    \item
           This view captures business cycles on which a specific API or microservice is provided, modified, updated, or expired and can be captured in tabular form, see Table~\ref{sys_when}.
            \begin{table}[H]
                \caption{ $\mu$-services and API life cycle}
                \label{sys_when}
                \centering
            \begin{tabular}{|c|c|c|c|} 
                 \hline
                  \rowcolor{Black}
                    \textbf{Business Cycles} &
                    \textbf{Provide} &
                	\textbf{Delete} &
                	\textbf{Update} \cr
                \hline    
                    \textbf{$\mu$-service$_{1} .. _{n}$} & 
                    business cycle& 
                    x& 
                    date\cr
                \hline 
                  \textbf{API$_{1} .. _{n}$} &
                    date& 
                    time& 
                    x\cr
                \hline
            \end{tabular}
            \end{table}
	   
\end{description}

\subsection{Technology Model (Builder's View)} 
\textquotedblleft Technology Model" is the description of microservices from a technology, data, and infrastructure perspective. The technology stack and the design decisions are captured in this viewpoint. These decisions play a major role in the CD and CI pipelines. For example, the configuration files (YAML files for containers and Kubernetes); the Software Development Kits (SDKs), and code samples are described in this \textit{viewpoint concern}. The data transfer mechanisms between microservices (XML, JSON, Protocol Buffers known as gRPC~\cite{grpc}, etc.) are also decided and captured in this \textit{viewpoint concern}. 

The following groups of stakeholders see the organization from this view: developers, programmers, designers, DevOps engineers, network engineers, SRE engineers, etc. Below is the detailed description of their \textit{viewpoint concerns} from each viewpoint perspective. 

\begin{description}
    \item [\textit{who (people):}] 
    \item
    This \textit{viewpoint concern} captures the detailed design decision of the APIs and microservices. The decision in which microservices will be system-level microservices and which ones will be user-level microservice is decided. Reverse proxy rules and considerations from the access perspective is also captured in this \textit{viewpoint concern}. 
    \item [\textit{what (data):}]
    \item
          This \textit{viewpoint concern} captures details of physical data elements to be used by each microservice and/or exposed by APIs. This is the point the builders provide details on how the data between the microservices will be interchanged (JSON, XML, or gPRC).
    \item [\textit{which (selection):}]
    \item
        This \textit{viewpoint concern} deals with the selections and prioritization related to physical data elements. Low-level design decisions for the design of microservices and APIs are captured in this \textit{viewpoint concern}. For example, the choice of containers available for a specific microservice and the priority given to a specific container (for its light-weightiness or availability in the private repositories). 
        
    \item [\textit{where (network):}]
    \item
        This \textit{viewpoint concern} captures the detailed design of network locations where microservices are deployed. This \textit{viewpoint concern} captures system networking requirements, such as Kubernetes clusters and the container networking requirements and needs, as shown in Table~\ref{builder_netwrok}.
            \begin{table}[H]
                \caption{ K8s clusters Networking}
                \label{builder_netwrok}
                \centering
            \begin{tabular}{|c|c|c|} 
                 \hline
                  \rowcolor{Black}
                    \textbf{POD$_{1} .. _{n}$} &
                    \textbf{POD CNI} &
                	\textbf{IP$_{1} .. _{n}$} \cr
                \hline    
                    \textbf{$\mu$-service$_{1} .. _{n}$} & 
                    IP& 
                    IPs \cr
                \hline 
                  \textbf{Container$_{1} .. _{n}$} &
                    IP& 
                    IPs\cr
                \hline
            \end{tabular}
            \end{table}
        
	\item [\textit{how (function):}]
    \item
     This \textit{viewpoint concern} captures system-level design decisions of the APIs. The technology stack for each microservice is captured and the design parameters of the APIs - the access, authorization, and authentication mechanisms, are captured in this \textit{viewpoint concern}. 

    \item [\textit{why (motivation):}]
    \item
        This \textit{viewpoint concern} captures business rules implementations for  microservices and APIs. 

    \item [\textit{when (time):}]
    \item
        
        This \textit{viewpoint concern} captures implementation details for the expiration and deletion of microservices and APIs at specific business cycles and events.
\end{description}

\subsection{Detailed Representations (Consumer's View)}
This viewpoint describes how external consumers see the enterprise from outside. The stakeholders in this group include the business owners of external enterprises, the architects, and the developers community outside the enterprise. These stakeholders' \textit{viewpoint concerns} are how to conduct business with the enterprise and how the systems-level integration will work.

The US-based technology giants, also known as  FANG (Facebook, Amazon, Netflix, Google) \cite{winseck2017geopolitical} offer SDKs and code examples that directly target this group of stakeholders. The terms like \textquotedblleft developers experience'' comes from providing this \textit{viewpoint concern}. Most of the industry leaders in technology, provide detailed descriptions of their APIs and issue SDKs of their services. Their target audiences are the designers, architects, and developers in consuming organizations (these audiences, in turn, steer the partner organizations' business leaders to buy services from the enterprise). In this \textit{viewpoint concern}, the needs of external stakeholders are captured and defined. These views and \textit{\textit{viewpoint concern}s} of stakeholders, enable other organizations to consume what the enterprise offers.

This view focuses on the detailed representation of the organizations' microservices and APIs - API documentation (Swagger, OpenAPI, GraphQL~\cite{graphql}); and the Microservices SDKs and code samples. These \textit{viewpoint concerns} can be captured in the form given in Table~\ref{consumers}. 

            \begin{table}[H]
                \caption{SDKs and Code Samples}
                \label{consumers}
                \centering
            \begin{tabular}{|c|c|c|c|} 
                 \hline
                  \rowcolor{Black}
                    \textbf{Consumers View} &
                    \textbf{SDK} &
                	\textbf{Code Samples} &
                	\textbf{Documents} \cr
                \hline    
                    \textbf{$\mu$-service$_{1} ... _{n}$} & 
                    details& 
                    details& 
                    details \cr
                \hline 
                  \textbf{API$_{1} ... _{n}$} &
                    details& 
                    details& 
                    details \cr
                \hline

            \end{tabular}
            \end{table}

\section{The Proposed Framework} \label{framework}

Based on the technologies described in Section~\ref{background} and the descriptions given in \textit{Methods} (Section~\ref{methods}), we organize \textit{viewpoint concerns} in tabular form, as shown in Table~\ref{table:w6h}. The columns in Table~\ref{table:w6h} represent the ordered viewpoints and the rows are  stakeholders' groups drawing  from  work by \citet{sultan2015ordering, sultan2018ordering}. Each cell in Table~\ref{table:w6h} contains  \textit{viewpoint concerns} of a given stakeholders' group from a specific viewpoint. The order and precedence of w6h viewpoints enables capturing holistic EA of the enterprise \cite{sultan2018ordering}.

\begin{table*}[!htbp]
\caption{The proposed framework: Rows are the views - how stakeholders view an organization, and the square brackets list key stakeholders who hold this view. Columns are stakeholders' viewpoints (in the order proposed by \protect \cite{sultan2015ordering, sultan2018ordering}). The cells are high-level stakeholders' \textit{viewpoint concerns} from a specific viewpoint, based on discussions in Sections~\ref{background} and \ref{methods}.}

\label{table:w6h} \centering
\begin{tabular}{|k{1.95 cm}|p{1.75cm}|p{1.75cm}|p{1.75 cm}|p{1.95cm}|p{1.95cm}|p{1.75cm}|p{1.75cm}|}
  \hline
			\cellcolor{white}&
			\textbf{(1)} &
			\textbf{(2)} &
		    \textbf{(3)} &
			\textbf{(4)} &
			\textbf{2,3/2,4 $\Rightarrow$ (5)} &
			\textbf{2,5 $\Rightarrow$ (6)} &
			\textbf{4,5 $\Rightarrow$ (7)}\cr
		\rowcolor{Black}
 
	    \textbf{Microservices ($\mu$-services)} 	&
			\textbf{People (who)} &
			\textbf{Data (what)} &
			\textbf{Selection (which)} &
			\textbf{Network (where)} &
			\textbf{Function (How) } &
			\textbf{Motivation (Why)} &
			\textbf{Time (when)} \cr
  \hline
  \textbf{Scope}(\emph{Ballpark View}): [Business Development Directors, CIOs, CFOs, and CSOs]& 
 \centering{list of business the enterprise may develop $\mu$-services for, and expose or consume APIs from} &
			\centering{List of things required for each $\mu$-service} and/or API &
			\centering{Which: prioritized list of organizations, $\mu$-services and APIs candidate for reuse} &
			\centering{List of locations for hosting $\mu$-services and related technologies} &
			\centering{List of business processes for design each $\mu$-service} &
			\centering{List of business goals/ strategies for each $\mu$-service and API} &
			\centering{List of cycles significant for $\mu$-service and APIs}  \cr
  \hline
  \textbf{Business Model}( \emph{Owners' View}): [Shareholders, Investors, Founders, Board of
Governors] &
        \centering{e.g., $\mu$-services workflow models} &
	    \centering{e.g., $\mu$-services semantic models} &
        \centering{ e.g., Which business process, data elements are required for $\mu$-services and APIs} &
        \centering{e.g., Datacenter locations where specific $\mu$-service reside} &
        \centering{e.g., Business taxonomy and linkage between business processes with $\mu$-services and APIs, as shown in Table~\ref{owner_function}} &
        \centering{e.g., Business goals related to $\mu$-services and APIs} &
        \centering{e.g., Business cycles when an API or $\mu$-service is provided, modified, updated, or expired.} \cr
  \hline
    \textbf{System Model}( \emph{Designer's View}): [enterprise architects, requirements engineers, privacy specialists, BCP planners]&
    	    \centering{e.g., Interface requirements for APIs and $\mu$-services} - Table~\ref{b2bb2c} &
		\centering{e.g., Logical data elements to be used by each $\mu$-service and exposed
        by APIs - Table~\ref{sys_data}} &
		\centering{e.g., Selection and prioritization of business processes/logical data elements required by API and $\mu$-services } &
		\centering{e.g., Distributed $\mu$-service and container architecture} - Table~\ref{sys_netwrok}, Gateway and reverse proxy rules - Table~\ref{sys_apigateway} &
		\centering{ e.g., $\mu$-service technology stack and API design mechanisms} - Table~\ref{sys_fun1} \& Table~\ref{sys_fun2}&
		\centering{e.g., $\mu$-service business rules}&
		\centering{e.g., Automated deletion / provisioning of $\mu$-services at pre-specified time} \cr 
  \hline
  
    \textbf{Technology Model}( \emph{Builder's View}): [DevOps engineers, programmers, network engineers, SRE engineers] &
    \centering e.g., $\mu$-services architecture (system vs user level $\mu$-services)and patterns &
    \centering  e.g., $\mu$-service physical data elements; data interface specifications (JSON, XML or gPRC)&
    \centering  e.g., Selection and prioritization of low-level design components, like containers, and pods etc. &
    \centering e.g., $\mu$-service and container deployment locations, locations of Kubernetes clusters and configuration files (.yaml) &
    \centering e.g., $\mu$-service technology stack, API access, authentication and authorization mechanisms &
    \centering e.g., Business rules implementations for $\mu$-services and APIs &
    \centering e.g., Zero-touch deployment scripts for auto provision/deletions of $\mu$-services \cr

  \hline    
    \textbf{Detailed Representations} (\emph{Consumer's View}):
        & 
    \multicolumn{7}{c|}{
    \makecell{\\SDKs, APIs and Code Samples; See Table~\ref{consumers}\\
     e.g., (curl -v -X GET https://api.sandbox.paypal.com/v1/payment-experience/web \\  -profiles/XP-8YTH-NNP3-WSVN-3C76 
     -H "Content-Type: application/json" \\
     -H "Authorization: Bearer Access-Token");
    }
    }
  \cr
  \hline

\end{tabular}
\end{table*}

\section{Discussion} \label{discussion}

In this publication, we described the fundamental components of microservices-based cloud-native application developments. We presented a structured methodology to capture and define stakeholders' \textit{viewpoint concerns} from microservices and APIs perspective. We used well-established viewpoints to gather the full-some views of the entire enterprise. This structured approach enables capturing and defining each stakeholder's needs. Once the full-some needs and \textit{viewpoint concerns} are captured, these enable engineering and analysis for strategic planning, alignment, reusability, and agility.

It is well established in the industry that there is a widening gap between EA and DevOps teams. The methodology we provided in this publication fills this gap by enabling DevOps teams to understand the stakeholders' \textit{viewpoint concerns}. At the same time, this enables enterprise architects to capture and define microservices-based DevOps developments in enterprise architecture constructs. 

The methods described in Section~\ref{methods} can be used to create a full blueprint of the entire enterprise from stakeholders' \textit{viewpoint concerns} perspective for microservices-based developments. This framework establishes a mind-map to start the architectural effort for microservice-based developments. The order of viewpoints enables asking and capturing the \textit{viewpoint concerns} in the right order without missing necessary and required information. 
With the advent of microservices-based DevOps developments, organizations are facing another challenge, known as BiModel IT\footnote{Gartner defines BiModal IT as "the practice of managing two separate, coherent modes of IT delivery, one is focused on stability and the other on agility"}. Most legacy applications may fulfill their functional requirements; these are not scalable or interoperable with the modern microservices-based architectures. The proposed framework enables dealing with these challenges by providing tools (the ordered stakeholders' \textit{viewpoints concerns}) to bridge the gap between these two inherently disjoint worlds.
In summary the proposed framework can help organizations:

\begin{itemize}
     \item Scale microservices based on stakeholders' needs;
     \item Reduce complexities of microservices based developments; 
    \item Link stakeholders' needs for increased sponsorship;
    \item Perform analysis for strategic planning and future growth;
    \item Promote and increase re-use of microservices;
    \item Bridge the gap between DevOps and EA teams;
    \item Create EA for microservices-based developments.
\end{itemize}

Disclaimer: The opinions expressed in this paper are those of the authors and not necessarily of the Government of Ontario. 
\section{Future Research} \label{future}

We intend to use graph theory to model and cluster stakeholders' \textit{viewpoint concerns} and microservices. For example, to retire weak microservices, which does not carry much business value, we may cluster a weighted graph of stakeholder's \textit{viewpoint concerns} and retire those. The framework presented in this publication enables such representations. 

We encourage EA practitioners to validate the proposed framework with data collected from their enterprises. These findings will enable novel work in stakeholders' viewpoints and microservice-based developments.


\bibliographystyle{ACM-Reference-Format}
\bibliography{main}

\end{document}